\documentclass[aps,prl,showpacs]{revtex4}

\usepackage{psfig} \usepackage{epsfig} \usepackage{amssymb}
\usepackage{amsmath}

\def\R{{\mathbf{R}}}

\def\eff{{\hbox{\small\it eff}}}
\begin{document}

\title{%
  Optimal Paths for Spatially Extended Metastable Systems Driven by
  Noise} 

\author{Weinan E$^1$, Weiqing Ren$^2$, and Eric Vanden-Eijnden$^2$}
\affiliation{$^1$Department of Mathematics and PACM,
  Princeton University, Princeton, NJ 08544\\
  $^2$Courant Institute, New York University, New York, NY 10012 }

\begin{abstract}
  The least action principle is exploited as a simulation tool to find
  the optimal dynamic path for spatially extended systems driven by a
  small noise.  Applications are presented for thermally activated
  switching of a spatially-extended bistable system as well as the
  switching dynamics of magnetic thin films.  The issue of nucleation
  versus propagation is discussed and the scaling for the number of
  nucleation events as a function of the terminal time and other
  material parameters is computed.
\end{abstract} 


\maketitle

\twocolumngrid

Thermal and other noises play a very important role in the behavior of
many systems. No matter how small the noise is, it always has a
non-trivial effect on long enough time intervals.  Thermal noise has
an especially important role for micro- or nano-devices since the
energy barrier between metastable states may come close to $k_BT$ in
such systems and it may in fact limit the smallest size of such
devices.  One example is found in magnetic recording industry where
the superparamagnetic limit, below which the data retention time
becomes too short for commercial purposes, is now considered a serious
limiting factor for the maximally achievable storage density
\cite{cow00}.  In order to better control such processes, it is
necessary to first obtain a detailed quantitative understanding of the
effect of the noises in such systems. For bistable systems that are
often used in storage and memory applications, issues such as the mean
switching time and the dynamics of the switching path have to be
addressed.  A major challenge is that most systems of interest are
spatially extended and switching cannot be described using only a few
degree of freedom. For instance, magnetization reversal in
micron-sized magnetic films proceeds by nucleation and domain wall
motion instead of coherent rotation\cite{kogrke00}.  Predictions of
reversal rates based stochastic coherent rotation models \cite{nee55}
(stochastic ordinary differential equations) are off by orders of
magnitude \cite{bra93}.  In these cases, one has to deal with the
added complexity of stochastic partial differential equations.
  
Traditionally the methods of choice for a quantitative understanding
of the effect of noise have been the Monte Carlo method or direct
simulation of the Langevin equation. For small level of noise,
however, these methods become prohibitively expensive. Wile the
maximal time step for Monte Carlo or Langevin simulations is still
restricted by the deterministic part of the dynamics, the effect of
the noise is significant only on time scales which are exponentially
large in the inverse of noise amplitude. This time scale is the
natural one for the noisy dynamics, and it is unaccessible by Monte
Carlo or Langevin simulations. For instance, for the magnetic problem
(of interest in magnetic recording), the times scale accessible to
Monte Carlo or Langevin is on the order of nanoseconds, and the time
scale of interest might be years.
  
In this paper we introduce a new numerical procedure to study
spatially extended systems driven by a small noise.  We are interested
in the effect of rare events over long periods of time.  Our method is
based on the theory of large deviations which provides a least action
principle for the most probable dynamic path \cite{frwe98}. Even tough
there has already been works using the least action principle for
ordinary differential equations (see e.g.  \cite{ber98}),
extending this in the context of spatially extended systems present
non-trivial numerical difficulties.  In the examples we consider,
namely thermally activated switching in a system with a bistable
potential modeled by the Ginzburg-Landau equations in one and two
dimensions and sub-micron-sized magnetic thin films, the numerical
computations are non-trivial because switching proceeds by nucleation
and domain wall motion which are both very localized in space and
time, and require using very fine grids. Also, the diffusion terms
present in these equations lead to very bad condition numbers, which
results in a loss of accuracy as well as an increase of the numbers of
iterations necessary for convergence.  We introduce an efficient
numerical method to overcome these problems which incorporates amongst
other techniques the use of suitable preconditioners, efficient
minimizations techniques such as the Quasi-Newton method, etc.

\begin{figure}
  \center \includegraphics[width=7cm]{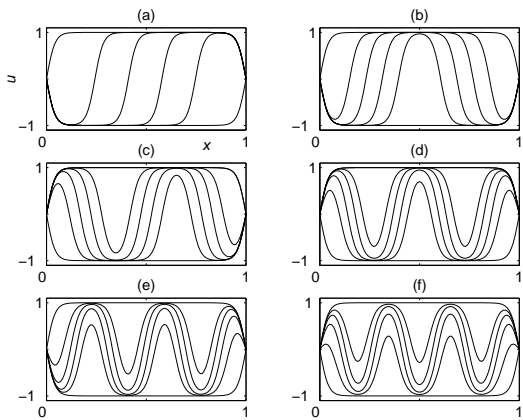}
  \caption{\label{fig:1}Snapshots of profiles of the minimizer
    $u$ during a switching from $u_+$ (top curve) to $u_-$ (bottom
    curve) at six equally spaced times on $[0,T]$ for different $T$ at
    fixed $\delta=0.03$: (a) $T=7$; (b) $T=2$; (c) $T=1$; (d) $T=0.8$;
    (e) $T=0.6 $; (f) $T=0.4 $. As $T$ decreases, the number of
    nucleations increases: one in (a), two in (b), (c), and (d), three
    in (e) and (f).  }
\end{figure}

Let us begin by considering the simplest continuum model describing
the spatio-temporal states of a bistable system, namely the
Ginzburg-Landau equation
\begin{equation}
  \label{eq:gl1d}
  u_t = \delta u_{xx} - \delta^{-1} V'(u),
\end{equation}
on the interval $[0,1]$ with the boundary conditions $u(0)=u(1)=0$. We
take $V$ to be a standard double-well potential, $V(u) =
\frac{1}{4}(1-u^2)^2$.  (\ref{eq:gl1d}) is expressed in appropriate
dimensionless variables in which $\delta$ is a small parameter which
indicates that the reaction term, $\delta^{-1} V'(u)$, is fast while
the diffusion is slow. (\ref{eq:gl1d}) can be considered as the
gradient flow associated with the energy
\begin{equation}
  \label{eq:energy1d}
  E[u] = \frac{1}{2}\int_0^1
  \left(\delta u_x^2 + 2\delta^{-1} V(u)\right) dx.
\end{equation}
The dynamics in (\ref{eq:gl1d}) has two stable equilibrium states,
$u_+$ and $u_-$, which minimize the energy (\ref{eq:energy1d}). When
$\delta$ is small, $u_\pm (x) =\pm1$ except at a thin boundary layer
of width $\delta$ at $x=0,1$ (see Fig.~\ref{fig:1}).

Now let us add to (\ref{eq:energy1d}) a small noise term modeling
thermal effects:
\begin{equation}
  \label{eq:gl1dn}
  u_t = \delta u_{xx} - \delta^{-1} V'(u) +\sqrt{\varepsilon}\, \eta ,
\end{equation}
where $\varepsilon$ is proportional to the temperature of the system
and $\eta$ is space-time Gaussian white noise with covariance
\begin{equation}
  \label{eq:cov}
  \langle\eta(x,t) \eta(y,s)\rangle = \delta(x-y) \delta(t-s).  
\end{equation}
The presence of the noise in (\ref{eq:gl1dn}) destroys the long-time
stability of the equilibria~$u_\pm$. For instance, if the initial
state is $u_+$, there is a finite probability that the system switches
to $u_-$ in any time interval $[0,T]$. To quantify the probability of
such a switching or, in fact, any other event, one can introduce an
action functional, $ S_{T}[u]$, such that for small $\varepsilon$, the
probability that the solution $u$ of (\ref{eq:gl1dn}) be close to a
given path $\varphi$ on $[0,T]$ can be estimated by
\cite{frwe98,fajo82}
\begin{equation}
  \label{eq:probest}
  \hbox{Prob}\{ u\approx \varphi\} \sim
  \exp(-\varepsilon^{-1} S_{T}[\varphi]).              
\end{equation}
The action functional corresponding to (\ref{eq:gl1dn}) is given
explicitly by
\begin{equation}
  \label{eq:action1d}
  S_{T}[u]  = \int_0^T \int_0^1
  \left(u_t - \delta  u_{xx} + \delta^{-1} V'(u)\right)^2 dx dt.
\end{equation}

The large deviation principle in (\ref{eq:probest}) allows us to
estimate the probability of various events associated with
(\ref{eq:gl1dn}) by constrained minimization of the action functional
(\ref{eq:action1d}). For instance, the probability $P_T$ that the
system switches from $u_+$ to $u_-$ before time $T$ satisfies
\begin{equation}
  \label{eq:ldestimate1d}
  \lim _{\varepsilon\to0} \varepsilon\ln P_T = -
  \min_{u}\left\{S_{0,T}[u]\right\},
\end{equation}
where the minimization in (\ref{eq:ldestimate1d}) is constrained by
the boundary conditions
\begin{equation}
  \label{eq:cons1d}
  u|_{x=0}=u|_{x=1}=0, \quad u|_{t=0}=u_+,\quad u|_{t=T}=u_-.
\end{equation}
Furthermore assume that the system switches before time $T$. Then the
minimizer of (\ref{eq:ldestimate1d}) gives the optimal switching path
between $u_+$ and $u_-$ during the time interval $[0,T]$, in the sense
that the probability that the system switches by another path is
exponentially smaller in $\varepsilon$.

\begin{figure}
  \center \includegraphics[width=6cm]{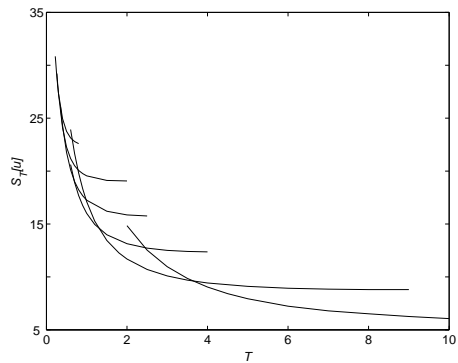}
  \caption{Values of the action as a function of
    switching time $T$ for the six minimizers shown in Fig.~1: the
    bottom curve corresponds to (a) in Fig.~1, while the top curve
    corresponds to (f).  Each switching path $u$ is a local minimizer
    of the action for all $T$: however, the glnobal minimizer changes
    with $T$.  } 
  \label{fig:2}
\end{figure}

To find the optimal path, we numerically minimize the action
functional, subject to the boundary condition in (\ref{eq:cons1d}).
We put down a numerical grid on the space-time domain
$(0,1)\times[0,T]$.  The action functional (\ref{eq:action1d}) is
discretized using finite difference formulas.  Numerical optimization
is implemented using a Quasi-Newton method, the BFGS method
\cite{noc99}. In order to speed up convergence, we used the operator
obtained from minimizing the linear part of (\ref{eq:action1d}),
$\int^T_0\int^1_0(u_t-\delta u_{xx})^2dxdt$, as the preconditioner.
Similar procedures were used for the other examples below and
numerical details are provided in \cite{ren}.

\begin{figure}
  \center \includegraphics[width=7cm]{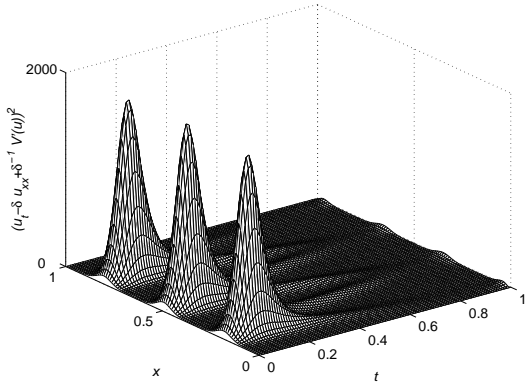}
  \caption{Space-time value of $\left(u_t - \delta \Delta u +
      \delta^{-1} V'(u)\right)^2$ for the minimizer (f) in
    Fig.~\ref{fig:1}. This can be interpreted as the minimal noise
    necessary for switching: the noise induces the nucleations then
    propagates the domain walls.  The peaks correspond to nucleations.}
  \label{fig:3}
\end{figure}

Figure~\ref{fig:1} shows the sequence of profiles of $u$ at different
times in $[0,T]$ for various values of $T$ at a fixed $\delta$
\cite{rem1}. The switching proceeds by nucleation followed by
propagation of domain walls. For large $T$, the switching proceeds by
propagating a domain wall from one boundary to the other. As $T$
becomes smaller, however, the number of nucleation events increase.
Such switching scenario with multi-nucleations are relevant at finite
temperature \cite{ren}.  In Figure~\ref{fig:2}, where we display the
values the action for the various local minimizers, plotted against
$T$.  Figure~\ref{fig:3} shows the space-time plot of $\left(u_t -
  \delta u_{xx} + \delta^{-1} V'(u)\right)^2$ during a switching
event.  This corresponds to the minimal (squared) noise necessary to
make a switching.

These results can be understood as follows. Consider
the critical points of the energy (\ref{eq:energy1d}), i.e. the
solutions of $0=\delta u_{xx} -\delta^{-1} V'(u)$ with $ u|_{x=0} =
u|_{x=1} =0$.  Besides $u_+$ and $u_-$, corresponding to the
minimizers of the energy, there are also saddle point configurations,
denoted by $u_S$, with an increasing number of domain walls.  For each
saddle point, there is a path joining $u_+$ to $u_-$ through that
saddle point.  Our result shows that, for large $T$, the switching
path crosses the saddle point configuration with minimum energy, i.e
the configuration with a single domain wall (see (a) in
Figure~\ref{fig:1}). As $T$ is decreased, one might naively think that
the switching will occur via the same path followed at a faster rate.
This intuition is wrong. For smaller $T$, the optimal switching path
crosses a saddle point configuration with increasing energy, i.e. more
nucleations and therefore more domain walls, giving rise to the
cascade of nucleation events seen in Figure~\ref{fig:1}.  The reason
is simply that both nucleation and domain wall motion are noise
induced. As $T$ decreases, at fixed number of nucleations the speed of
propagation of the domain wall must increase in order to achieve
complete switching during the allowed time $T$. This is energy
consuming and, at certain critical values of $T$ it becomes more
favorable to make an additional nucleation.

The same type of cascade in the number of nucleations is also observed
if $T$ is kept fixed but $\delta$ is decreased, because the cost of
domain-wall motion increases as $\delta$ decreases, whereas the cost
of a nucleation stays the same. In fact, a simple scaling argument
gives $n_\star= C (\delta T)^{-1/2}$, for small $\delta$ and $T$,
where $n_\star$ is the number of nucleations in the optimal switching
path, and $C$ is a numerical constant. Furthermore, one has $\lim
_{\varepsilon\to0} \varepsilon \ln P_T = -C' (\delta T)^{-1/2}$, which
gives the envelop of the curves in Fig.~\ref{fig:2}.  The detailed
analysis of the cascade process of nucleations will be presented
elsewhere.

\begin{figure}
  \center \includegraphics[width=6cm,angle=270]{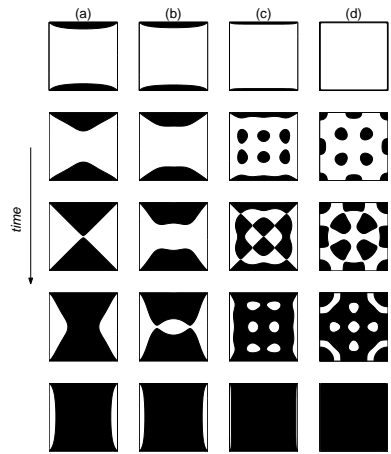}
  \caption{Snapshots of profiles of the
    minimizer $u$ during a switching from $u_+$ (top figure in one
    column) to $u_-$ (bottom figure in one column) at five equally
    spaced times on $[0,T]$ for $T=1$ and different $\delta$ and
    different boundary conditions: (a)--(c) correspond to case (i),
    (d) to case (ii). In (a) $\delta= 0.04$; (b) $\delta= 0.03$; (c)
    $\delta= 0.01$; (d) $\delta= 0.01$. The gray-scale is from white
    for $u=-1$ to black for $u=+1$.}
  \label{fig:4}
\end{figure}

As an example in two dimensions, we minimize 
\begin{equation}
  \label{eq:actiondd}
  S_{T}[u]  = \int_0^T \int_\Omega
  \left(u_t - \delta \Delta u + \delta^{-1} V'(u)\right)^2 d^2x dt,
\end{equation}
where $\Omega$ is the unit square, $\Omega=[0,1]\times[0,1]$.  We will
present results with two different boundary conditions: (i) $u=1$ at
$x=0$ and $x=1$, $u=-1$ at $y=0$ and $y=1$, and (ii) $u=0$ on the edge
of the square.  In both cases there are two global minimizers of the
energy. One minimizer, $u_+$, is close to $u=1$ except for the
boundary layers at $y=0$ and $y=1$ (case (i)) or at the edge of square
(case (ii)). The other minimizer, $u_-$, is close to $u=-1$ except for
the boundary layers at $x=0$ and $x=1$ (case (i)) or at the edge of
square(case (ii)). In the absence of noise, both minimizers are stable
equilibrium states.

In Figure~\ref{fig:4}, we show the time sequences of the switching
process for different values of $\delta$ at fixed $T$.  The overall
trend is consistent with what was found in the previous example,
namely there are more and more nucleation events as $\delta, T \to 0$.

Finally let us consider thermally activated switching of a magnetic
thin film, modeled by the Landau-Lifshitz-Gilbert equation, which
after suitable nondimensionalization, reads
\begin{equation}
  \label{eq:LL}
  m_t = f(m) := - m \times h_\eff - \alpha m \times (m \times  h_\eff).
\end{equation}
Here $m$ is the magnetization distribution (3d vector) which satisfies
$|m|=1$, $h_\eff$ is the effective local field, given in terms of the
free energy by $ h_\eff = -\delta E[m] /\delta m,$ where
\begin{displaymath}
  E[m] = \int_\Omega |\nabla m|^2 d^3x+  \int _\Omega \phi(m) d^3x+
  \int_{\R^3} |\nabla u|^2 d^3x.
\end{displaymath}
Here the three terms represent respectively energies due to exchange,
anisotropy, and stray field.  $\Omega$ is the region occupied by the
magnetic thin film. The potential $u$, defined everywhere in space,
solves $\hbox{div} \left(-\nabla u + m\right) =0$, where $m$ is
extended as $0$ outside the sample.

\begin{figure}
  \center \includegraphics[width=4.5cm]{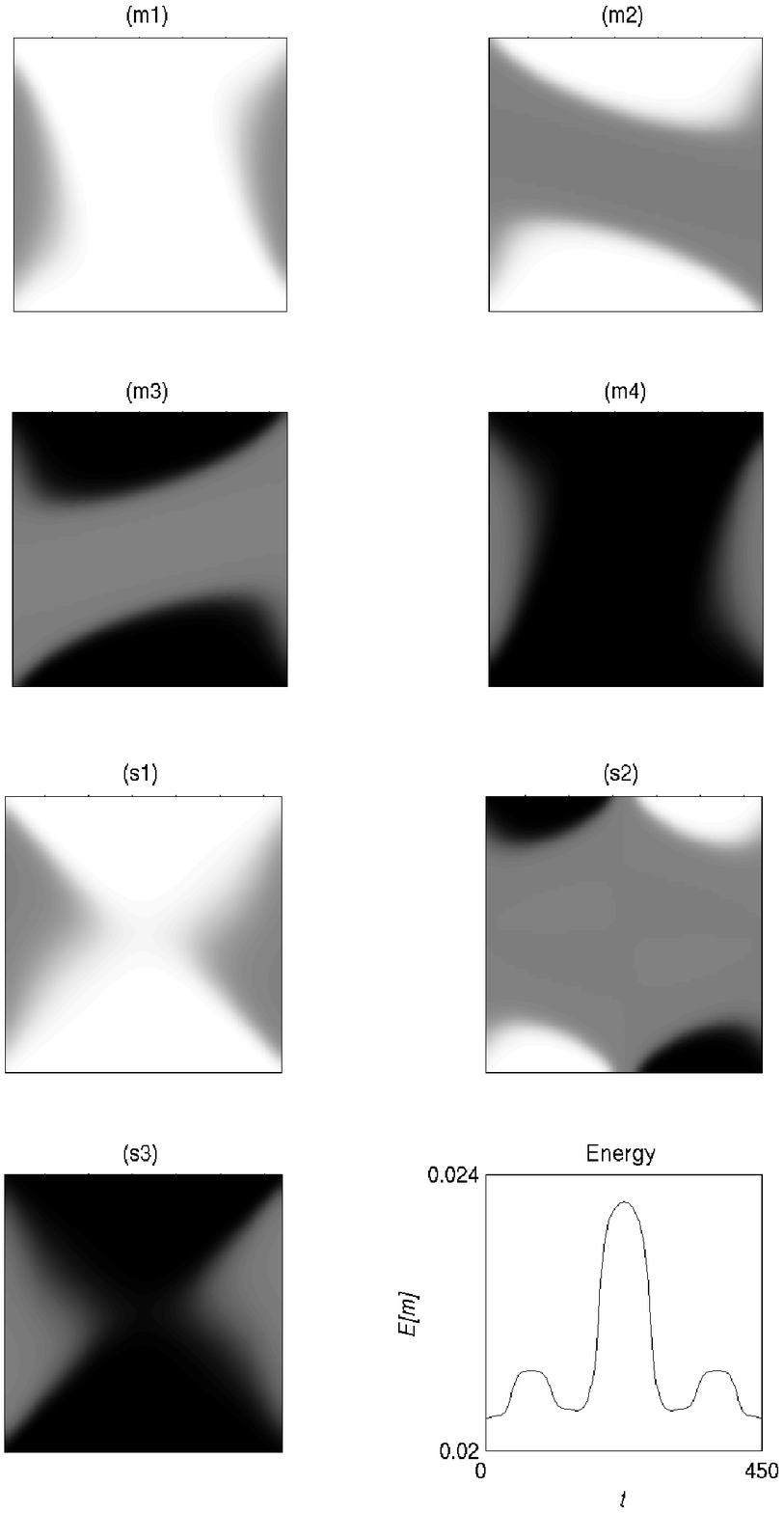}
  \caption{
    Minima and saddle points configurations crossed by the optimal
    switching path from (m1) to (m4); the energy of the magnetization
    during the switching shows that two intermediate minima, (m2) and
    (m3), and three saddle configurations, (s1), (s2), and (s3), are
    crossed during the switching. The out-of-plane component of the
    magnetization vector $m$ is very small (less than $10^{-2}$)during
    the switching and we only plot its in-plane component: it points
    to the right in the white areas, to the top in the gray areas, and
    to the left in the black areas.  }
  \label{fig:5}
\end{figure}

The first term at the right hand side of (\ref{eq:LL}) describes
precession of $m$ around $h_\eff$; the second term, which can be
written as $- \alpha m \times (m \times h_\eff)= \alpha ( h_\eff -
(m\cdot h_\eff) m)$, is a damping term whose strength is measured by
the parameter $\alpha$. We choose physical parameters corresponding to
permalloy, and consider a
thin square domain of size $200nm \times 200nm \times 10nm$.

The dynamics in (\ref{eq:LL}) has equilibrium states corresponding to
the minima of the magnetic energy, $E[m]$. Thermal noise effects can
be represented by modifying $h_\eff$ as $h_\eff + \sqrt{\varepsilon}\,
\eta$, where $\eta$ is a vector white-noise, and will eventually
switch the system from one minimum to any other one. In
Fig.~\ref{fig:5}, (m1) and (m4) show two minima where the
magnetization points mostly to the right (white color) or to the left
(black color), respectively.  For small noise, the optimal path for a
switching from (m1) to (m4) during the time interval $[0,T]$ is
obtained by minimizing the action functional corresponding to
(\ref{eq:LL}):
\begin{equation}
  \label{eq:actionLL}
  S_T[m]= \int_0^T \int_{\Omega} \bigl|m_t -f(m)\bigr|^2 d^3xdt.
\end{equation}
The global minimizer of (\ref{eq:actionLL}) on a long time interval,
$T=450$, is presented in Fig.~\ref{fig:5}. Looking at the energy
$E[m]$ during the switching one sees that two intermediate minima,
(m2) and (m3), and three saddle configurations, (s1), (s2), and (s3),
are crossed during the switching.  The first step of the switching
from (m1) to (m2) through (s1) rotates the interior magnetization by
$90^o$.  The second step from (m2) to (m3) through (s2), which is the
most expensive step, switches magnetization at the top and bottom end
domains. Finally, the third step from (m3) to (m4) through (s3) is
similar to the first one.

In conclusion, we have shown that the least action principle can be
turned into an efficient numerical procedure for finding the optimal
dynamic path in spatially- extended systems driven by small noise, and
we have presented applications to bistable systems modeled by the
Ginzburg-Landau equation and the magnetic thin films.  Our numerical
method circumvents the difficulty of having to compute too many time
steps in order to observe the relevant events. It is general and can
be applied to a variety of problems in physics, chemistry as well as
biology.

Weinan E is partially supported by NSF through a Presidential Faculty
Fellowship.  Eric Vanden-Eijnden is partially supported by NSF Grant
DMS-9510356.

\end{document}